# A Historical Study on Development of Optometry

Dal-Young Kim[1]

Department of Optometry, Seoul Tech, Seoul 139-743, Republic of Korea

Development of optometry in western countries was studied on a viewpoint of the history of science. It was revealed that optometry had been formed on the basis of optics, a branch of physics, to which biomedical study was added. Optometry can be defined as a pioneering interdisciplinary field of study and as biomedical physics of the 19th century. It is pointed out that population-geographic factors have affected on the development of optometry in the U. S. A. and Australia. European development model is suggested for a strategy for the next-generation development of Korean optometry, rather than American model.

**Keywords:** Ophthalmic Optics, Optometry, History of Science

[1] Department of Optometry, Seoul National University of Science and Technology (Seoul Tech)
172 Gongreung 2-dong, Nowon-gu, Seoul 139-743, Republic of Korea
E-mail address: dykim@seoultech.ac.kr

# 안경광학의 발전 과정에 관한 과학사적 고찰

김달영[2]

서울과학기술대학교 안경광학과

서구 선진국의 검안학/안경광학의 발전과정을 과학사적 측면에서 고찰하였다. 검안학/안경광학은 물리학의 한 분야인 광학을 기반으로 생물의학 연구가 결합하여 형성되었음을 밝히고, 선구적인 학제간 학문이며 19세기의 생물물리학이라고 규정될 수 있음을 보였다. 미국과 호주의 검안학 발전 과정에 있어 인구지리학적 요인이 중요한 영향을 미쳤음을 지적하고, 현재 한국 안경광학계가 당면하고 있는 문제점들을 해결할 수 있는 발전 방향으로 영미식 모델보다는 대륙식 모델이 더 적합함을 제안하였다.

**주제어:** 안경광학, 검안학, 과학사

[2] 우편번호 139-743 서울시 노원구 공릉2동 172번지 서울과학기술대학교 안경광학과
전자우편주소: dykim@seoultech.ac.kr

# 1. 서론

검안학(檢眼學) 또는 안경광학(眼鏡光學)은 영어의 optometry를 번역한 용어이며, '사람의 눈과 관련하여 안보건, 눈의 구조, 시력, 시각과 관련된 눈과 뇌의 시스템, 시각정보처리 등에 관하여 연구하는 학문'이라고 정의할 수 있다 (김재도, 이익한 2008). 조금 더 구체적으로 표현한다면, 시각기능의 메커니즘에 대하여 연구하고 그 이상 증상을 광학적인 수단으로 교정하는 방법을 개발하는 실용학문으로서, 한국, 일본, 유럽의 안경사(optician)나 미국 또는 영연방 국가들에서 검안사(optometrist)가 되기 위해서는 반드시 공부하여야 하는 필수전공으로 지정되어 있다.

학문의 명칭이나, 검안사 또는 안경사의 면허 취득 과정에서 필수전공으로 지정되어 있다는 사실에서 알 수 있듯이, 검안학/안경광학은 기본적으로 안경이나 콘택트렌즈를 사용하여 사람의 시력이상 또는 굴절이상을 교정하는 것을 목적으로 하고 있다. 시력이상 또는 굴절이상은 사람의 눈에서 빈도 높게 발견되는 증상이며, 사람이 입수하는 정보의 절반 이상이 시각 정보라는 사실이나 IT 정보화 시대를 맞이하여 시각 정보의 비중이 갈수록 증대되고 있다는 사실을 고려할 때, 건강한 시력의 중요성은 아무리 강조해도 지나치지 않다고 할 수 있다.

대부분의 실용학문이 그렇듯이, 안경광학은 안경과 콘탠트렌즈, 그리고 시력이상과 교정이라는 목표와 대상을 지향하며 형성된 학문이며, 자연과학, 철학, 사회과학처럼 방법론에 의해서 확립된 학문 분야가 아니다. 그렇기 때문에 안경이나 시력과 관련된 여타 다른 학문 분야의 성과나 방법론이 다양하게 안경광학에 수용되어 있다. 국가에 따라 검안사/안경사의 업무 범위가 조금씩 다르기 때문에, 안경광학의 교육과 연구 내용도 다양하다. 한국과 미국을 기준으로 보았을 때, 안경광학의 주요 세부분야는 크게 시광학(視光學, visual optics), 안과학(眼科學, ophthalmology), 안경학(眼鏡學)으로 구분할 수 있다. 시광학은 주로 물리학 또는 물리학의 세부 분야 가운데 하나인 광학에 속하는 영역이고, 안과학은 의학 또는 생명과학의 영역에 속하며, 안경학은 안경 자체나 안경원에 대한 학문으로서 실무경험의 측면이 강하다. 따라서 안경광학은 안경과 콘택트렌즈와 연관된 모든 주제를 아우르는 종합학문의 성격을 가지고 있다 할 수 있겠다.

안경광학은 19세기 말엽부터 20세기 초반에 걸쳐 실용적인 필요에 의하여 형성된 학문이다. 미국의 학계에서는 optometry 학문이 시작된 시기를 1890년대로 보고 있다 (Goss 2003). 그러나 차후에 논의되듯이, 실질적으로 안경광학이 본격적으로 발전하여 엄격한 의미에서 독립적

인 학문으로 형성된 시기는 20세기 초중반 정도로 보는 것이 타당하다 하겠다. 안경광학은 수학이나 물리학 또는 철학처럼 오랜 역사를 가진 지적 탐구의 결과가 아니고, 경영학이나 행정학과 유사하게 20세기 들어서 필요에 따라 학제간 전공으로 새롭게 발생된 학문인 것이다.

이러한 안경광학의 발전과정을 연구하는 것은 두 가지 측면에서 의의를 갖는다고 할 수 있다. 첫 번째 의의는 안경광학이라는 독특한 학문이 과학의 발전이라는 커다란 틀 속에서 어떤 위치를 차지하는가라는 과학사학적인 질문에 답하는 것이다. 뒤에서 논의되듯이, 안경광학이라는 학문 분야의 형성 과정은 과학사적으로 흥미있는 몇가지 특징을 가지고 있기 때문이다. 두 번째 의의는 흔히 일컬어지는 것처럼 '역사로부터 배운다'는 교훈을 실천할 수 있다는 점이다. 현재 한국의 안경계와 안경광학계는 여러 가지 난제를 가지고 있다. 검안사 제도의 도입 여부 및 그와 연관된 안경광학의 정체성 문제에서부터 시작하여, 안경사 교육의 교과 과정과 범위 문제를 거쳐, 장기 불황에 따른 안경 사업의 침체 문제에 이르기까지 매우 다양하다. 미국을 중심으로 하는 서구 선진국의 안경광학 발전 과정을 연구함으로써 현재 한국의 안경광학계가 직면하고 있는 여러 문제에 대한 답을 얻을 수 있으리라는 기대가 본 연구의 주요한 동기 가운데 하나이다.

## 2. 검안학/안경광학 학문 분야 형성 이전 시광학의 발전 과정

빛은 시각에 의하여 사람이 가장 기본적으로 인지할 수 있는 외부 자극이기 때문에, 빛에 대한 탐구는 아주 오랜 옛날부터 시작되었으며, 광학이라는 학문의 기본이 되는 몇 가지 발견이 근세 이전에 이미 성취되었다. 플라톤, 프톨레마이오스, 중세 아랍과 유럽의 광학 연구 등이 그것이다. 이러한 빛에 대한 경험적 연구의 축적은 과학혁명기에 케플러, 호이헨스, 데카르트, 뉴턴, 로버트 후크 등 여러 물리학자들의 관심을 끌었으며 그 결과 19 세기 초반에 이르러서는 광학이 물리학의 한 분야로 확고하게 인식되게 되었다.

물리학 전체의 학문 구조를 놓고 보았을 때, 광학은 약간 독특한 면을 가진다. 물리학은 기본적으로 물질 체계 전체를 이해하려는 학문이고, 연구의 대상에 따라 카테고리가 정해지는 것이 아니라 물리학 특유의 연구 방법론이 그 특성을 규정하는 학문이다. 물리학의 4대 기초 과목이라고 일컬어지는 고전역학, 전자기학, 양자역학, 통계역학도 모두 연구방법론에 의하여 구분된 것이며 특정대상을 연구하는 세부 분야가 아니다. (전자기학의 경우 전기와 자기라는 대상이라고 생각될 수도 있으나, 이 대상이 독립적인 과목이 된 이유 자체가 전자기에 적용되는 연구방법론이 역학이나 통계와 달랐기 때문이라고 할 수 있다.) 그에 반해서 광학은 '빛'이라는

연구대상이 확고하게 규정되어 있는 독특한 물리학 분야이며, 나중에 물리학 연구가 진전되면서 빛이 곧 전자기파임이 밝혀져서 전자기학과의 연관성이 밝혀졌지만, 19세기 중반까지만 해도 빛은 물리학의 기본인 역학과는 거의 상관이 없는 동떨어진 연구주제였다. 과학혁명기 이래로 광학이 물리학의 한 분야로 확고하게 인정받은 이유는, 아마도 당시의 저명한 물리학자들이 광학 연구를 병행했기 때문이고, 다른 물리 분야처럼 수학적인 분석이 용이했기 때문이라고 짐작된다. 빛에 대한 연구는 19세기 중반까지 역학이나 다른 물리학 분야와 연관된 사고틀(frame)으로 이해되는 주제가 아니었다.

이처럼 물리학의 세부 전공이면서도 전체적인 물리학 체계에서 동떨어진 면을 가지고 있던 광학 연구는, 전자기파에 대한 연구가 발전하면서 전자기 현상과 통합되어 물리학의 일부로 편입 되어가는 한편으로, 시각에 대한 생물학적 심리학적 연구와 결부되어 시광학(視光學, visual optics) 분야를 형성하면서 발전하는 양상을 보이기 시작했다. 애초에 인간이 빛에 대하여 관심을 가지게 된 이유가 시각이 존재하기 때문이므로, 광학 연구가 인간의 시각에 대한 연구와 연관되는 것은 매우 자연스러운 일이었다. 케플러, 뉴턴, 토머스 영 (Atchison, Charman 2010)과 같은 물리학자들이 각각 근시, 원시, 난시에 대한 이론을 확립하였다는 사실에서 시광학이 물리학/광학으로부터 파생된 학문이라는 사실을 쉽게 알 수 있다 (케플러의 경우는 그 자신이 심한 근시 환자이기도 했다).

시광학이라는 학문의 분화 및 형성에 결정적으로 기여한 사람은 그 또한 물리학자였던 헬름홀츠라고 할 수 있다. 물리학자가 되기 전에 의학 교육을 받았고 의대 교수직까지 지냈던 독특한 이력의 소유자인 헬름홀츠는 자신의 교육배경을 살려서 광학에 대한 물리학적 지식을 안과학에 접목시켜 안광학(眼光學, ophthalmic optics)이라는 새로운 학문분야를 창출하였으며 눈의 수정체의 조절(accommodation) 작용에 대한 이론을 수립하여 해부학, 생리학, 광학(물리학)의 지식을 동시에 구사하는 흥미로운 학제간 연구주제를 후학들에게 제시하였다. 헬름홀츠는 조절 작용의 이론에만 그치지 않고 색각(色覺, color vision)이나 공간시각(spatial vision)에 대한 연구에도 주요한 기여를 하면서, 당시 권위있는 심리학자였던 헤링(E. Hering)과의 논쟁을 통하여 시각에 대한 연구가 심리학과 연관되는 계기를 만들기도 하였다.

이러한 시광학의 발전 과정을 과학사적으로 음미하여 보면, 한 가지 흥미로운 점을 지적할 수 있다. 그것은 광학을 사람의 눈에 적용하여 발전된 시광학이 현대의 생물물리학(biophysics) 또는 의학물리학(medical physics)과 유사한 의의를 갖는 19세기의 선구적인 학제간 연구라는 사실이다. 생물물리학이나 의학물리학의 정확한 정의에 관해서는 다양한 의견이 있으나, 물리학에서의 연구 성과를 생물학이나 의학 연구에 적용하는 학제간 연구 분야라는 점

에서는 대체로 의견이 일치하고 있다. 19세기의 시광학은 이러한 생물물리학/의학물리학의 의미에 정확하게 해당된다고 할 수 있으며, 20세기 중반에야 형성되기 시작한 현대적인 생물물리학/의학물리학에 비하여 한 세기 정도 빠르게 먼저 물리학의 연구 결과를 인체에 적용하여 의미있는 통찰을 창출해낸 19세기의 학제간 연구 분야라고 볼 수 있다. 21세기 들어와 학문의 통섭과 학제간 연구에 대한 강조가 두드러지고 있는 현실에서, 기존의 학제간 연구분야들보다 한 세기 일찍 학제간 연구라는 방법론과 생물학/의학에의 물리학 응용을 구현한 시광학의 과학사적 의의는 주의 깊게 연구할 가치가 있다고 사료된다.

## 3. 검안학/안경광학 학문 분야의 형성 과정

이와 같은 과정을 거쳐 시광학은 물리학의 하부 전공인 광학의 세부 전공으로 확립되며, 광학의 의학적 응용이라는 독특한 특성을 나타내며 물리학이나 광학으로부터 분화하여 독립하기 시작했다. 안경광학이라는 학문이 독자적으로 형성되는 데에는 시광학의 기여가 가장 컸다고 할 수 있다. 시광학을 눈에 적용함으로써 과학적 이론에 기반하여 시력 이상을 합리적이고 체계적으로 렌즈를 이용하여 교정할 수 있게 되었기 때문이다. 검안학/안경광학의 역사를 탐구하는 연구자들은 검안학/안경광학과 안과학의 가장 큰 차이점으로, 안경광학은 눈의 '광학'을 중심으로 우선(굴절이상교정)하는데 반하여, 안과학은 눈의 '의학'을 우선시(안질환의 진료)한다는 점을 꼽고 있다 (Goss 2003). 따라서 검안학/안경광학이라는 학문분야가 독자적인 학문 분야로 진화하는데 있어 시광학의 역할이 가장 중요했음을 알 수 있다.

물론 단순히 시광학이 검안학/안경광학으로 단선적으로 진화한 것은 아니다. 안경광학은 시광학의 뼈대 위에 의학(안과학)을 접목시킴으로써, 연구대상만 생체이지 방법론은 완전히 물리학의 틀을 따르던 시광학과 달리, 사람의 신체(즉 시력)을 교정하는 보건의료 분야의 학문으로 발전한 것이다. 다시 말해서, 안경광학 분야의 형성에는 앞서 언급된 시광학적 기반에 더하여 의학의 일부인 안과학도 주요한 역할을 했음을 고려해야 한다. 안과학은 주지하다시피 의학의 한 분과로서 오랜 전통을 가지고 있으며, 눈과 연관되는 생리학, 병리학, 조직학 등의 지식이 검안학/안경광학 내에서 상당한 부분을 차지하고 있다. 따라서 안경광학은 아주 단순하게 이해한다면 시광학의 전통 위에 안과학이 접목되면서 탄생한 분야라고 할 수 있다.

안경광학/검안학이 단순한 시광학의 확대가 아니라는 사실은 그 명칭으로부터도 알 수 있다. 검안학(optometry)라는 용어가 정착되기 이전까지 안경광학 분야는 생리광학(physiological optics)라는 명칭으로 지칭되었으며, 시광학과는 명칭을 달리 하여 독립된 학문 분야로서의 정

체성을 분명히 하였다. 생리광학이라는 명칭은 현재도 일부 검안학/안경광학 교육 조직의 전공 명칭으로 남아 있다 (University of Houston 2012).

앞 절에서 언급되었던 시광학과는 또 다른 의미에서, 검안학/안경광학은 21세기에 주목받고 있는 학제간 신학문의 선구적인 예제라고 할 수 있다. 시광학의 경우에는 광학의 방법론을 적용시키는 대상을 생체(눈)으로 확장시킴으로써 광학의 새로운 세부 분야를 확장한 것이지만, 검안학/안경광학은 시광학에 안과학이 접목됨으로써 광학이나 안과학과는 전혀 다른 새롭고 독립적인 새로운 학문 분야가 발생된 것이기 때문이다. 즉 기존 학문 내의 세부 분야가 아닌, 학제간 연구를 통한 독립적인 학문의 생성이라는 점에서 검안학/안경광학의 형성 과정은 시광학의 분화와는 다르다고 할 수 있다.

검안학/안경광학의 형성은 단순히 시광학과 안과학의 결합으로만 해석할 수는 없다. 장식용으로나마 렌즈가 이용되기 시작한 것은 고대 이집트에서 부터였으며, 짧게 잡더라도 13세기에는 안경이 제작되고 있었다. 따라서 안경 그 자체는 19세기 말 부터 20세기 초에 검안학/안경광학이라는 학문 분야가 형성되기 몇 세기 전에 이미 존재하였음을 알 수 있다. 체계적으로 교육되고 연구되는 정식 학문 분야는 아니었으나, 안경을 제작하고 개발하는 기술인 안경학은 검안학/안경광학의 형성 이전부터 이미 존재하고 있었던 것이다. 철학자 스피노자의 직업이 안경용 렌즈 연마사였다는 사실은 널리 알려져 있다. 다시 말해서, 안경이라는 직업 분야는 학문 분야가 형성되기 수백 년 전부터 이미 존재하고 있었고, 안과학이 첨가된 시광학 연구의 전통이 기존의 안경 직업 분야와 결합하면서 검안학/안경광학 학문 분야가 형성된 것이다. 이러한 검안학/안경광학의 성립 과정에서는 공학의 형성 과정과 유사한 점을 찾아볼 수 있다.

경험에 바탕한 실용 기술이라는 의미에서의 공학 또는 엔지니어링은 기원 이전부터 존재하였고, 근세 이후에는 군의 공병 학교 등에서 체계적으로 교육되고 연구되어 왔다. 이러한 실용 기술이 먼저 존재하였고, 그에 물리학 등에서 얻어진 과학적 분석법이 결합되고 대학과 같은 고등교육기관에서의 교육이 이루어지면서 현대와 같은 형태의 공학이라는 학문 분야가 탄생한 것이다. 실용기술이 먼저 존재하고 그에 관한 과학적인 연구가 더해지면서 학문 분야가 확립되었다는 점에서 안경광학은 공학과 유사한 형성 과정을 밟았다고 할 수도 있다.

미국의 경우를 기준으로 할 때, 19세기까지 안경의 처방과 제작은 귀금속상, 시계공, 약행상인, 무자격 떠돌이 의사 등에 의해서 이루어지는 경우가 많았으며 대도시보다는 교통이 불편한 지방일수록 이런 현상이 더 심각했던 것으로 여겨진다 (Fiorillo 2010). 전문성은 높지 않았지만 안경만을 전문으로 하는 안경사도 존재하였으나, 오늘날과 같은 개업안경사의 개념이 아니고 행상에 가깝게 간단한 장비만을 소지한 채 떠도는 경우가 많았다. 이 가운데 특히 중요한

것은 귀금속상(jeweller)이다. 귀금속상은 업무상 보석 감정에 쓰이는 광학적인 지식과 보석에 부가되는 금속장식 부분과 관련한 조제 가공 기술을 갖추고 있었는데, 광학 지식과 금속조제가공 기술은 안경사의 실무에 필수적인 요소이기 때문에 귀금속상이 안경사를 겸하기가 용이했다. 그 때문에 안경사-귀금속상의 겸업이 매우 흔했으며 (McBroom 1998), 한국에서도 안경사 제도가 확립되기 이전까지는 귀금속상이 안경 업무를 겸하는 경우가 많았다. 한국에서는 오늘날까지도 드물게나마 귀금속상과 안경원을 겸업하는 경우를 볼 수 있다.

미국에서 안경사에 대한 교육은 19세기 후반부터 이루어지고 있었지만 오늘날의 기준으로 보면 직업 학원 수준에도 미치기 어려운 수개월간의 실무 교육 정도 수준이 대부분이었으며 국가나 주가 관리하는 어떤 형태의 공인된 면허 제도도 존재하지 않았다 (Fiorillo 2010). 미국 최초의 검안학 교육 기관은 1872년 창립된 중서부의 Illinois College of Optometry이며 1894년에 동부의 New England College of Optometry (NEWENCO), 1904년에 서부의 Southern California College of Optometry가 설립되었다 (Goss 2003). 그러나 19세기 동안 이러한 교육기관들은 실무를 교육하는 직업 학원 수준에 머물렀던 것으로 보이며, 실질적으로 체계적이고 본격적인 교육이 이루어진 것은 20세기에 들어서 정식으로 대학(university)에 검안학 교육과정이 설치된 이후라고 해석되고 있다.

## 4. 영어권 국가에서 검안사 직업 및 검안학이 발전된 이유

위에서 언급된 것과 같이 시광학, 안과학, 기존의 안경기술이 융합하는 과정을 거쳐, 20세기 초반부터 검안학/안경광학은 안과학이나 시광학과는 별도의 독립적인 학문으로 독립하기 시작했다. 검안학/안경광학은 독립적인 학문분야일 뿐만 아니라, 검안사 또는 안경사라는 전문 직업군의 실무지식을 교육하고 연구하는 실용과학이기도 하다. 따라서 학문의 특징이 해당 직업의 업무 영역이나 관련 제도의 영향을 크게 받는다. 여기서 흥미로운 점은 검안학/안경광학과 관련된 직업 제도가 국가별로 큰 차이를 보인다는 사실이다.

검안사/안경사 제도의 형태를 대략적으로 구분하면 영미식과 대륙식으로 구분할 수 있다. 대륙식은 유럽 대륙의 프랑스, 독일 등에서 주로 적용하고 있는 제도로써, 안보건(眼保健)에 관련된 전문직업이 안과의사와 안경사로 2원화 되어 있으며, 렌즈를 이용한 시력 교정 업무는 주로 안경사가, 눈의 질환 치료는 주로 안과의사가 담당하는 시스템이다. 이러한 제도를 일본이 모방하여 현재 일본도 안과의사-안경사 방식의 2원화된 안보건 체제를 가지고 있으며, 한국도 일본의 영향을 받아 동일한 안보건 체제를 운용하고 있다. 체제는 동일하더라도 전문직군의 자격 제도는 나라마다 약간씩 다르다. 독일과 한국은 국가가 안경사 면허를 관리하며 규정

된 교육을 받고 국가면허시험에 합격해야 하는데 반해서, 프랑스와 일본은 안경사의 국가면허 제도가 없고 민간 차원에서 안경사의 질적 수준을 관리하고 있다 (McGrath 2003).

영미식은 미국과 영어권 국가인 미국, 캐나다, 호주, 뉴질랜드 및 영국의 과거 식민지였던 몇몇 나라에서 적용되고 있는 제도로써, 안보건 전문직군이 안과의사-검안사-조제가공안경사로 3원화되어 있다. 영미식 제도에서 조제가공안경사는 직무영역이 제한되어 있어서, 눈에 대한 광학적 검사를 할 수 없고 안과의사나 검안사가 발급한 처방전에 따라 안경 또는 콘택트렌즈를 조제, 가공, 판매하는 일만 할 수 있다. 업무 범위가 좁기 때문에 영미권의 조제가공안경사는 교육기간이 수개월에서 1년 내외로 짧고, 면허시험도 그다지 난이도가 높지 않다. 두 종류의 안경사를 구분하여, 눈에 대한 광학검사를 수행할 수 있는 대륙식 안경사를 ophthalmic optician이라 지칭하고, 검사를 수행할 수 없는 영미식 조제가공안경사는 dispensing optician이라고 다르게 호칭하기도 한다. 검안사는 한국에 없는 영미식 안보건 체제의 독특한 전문직업인데, 안경 또는 콘택트렌즈의 처방을 위한 검안 및 시력검사 업무를 기본으로 하고, 눈 수술과 같은 고급 업무를 제외한 기초적 안과 진료 업무도 함께 수행할 수 있는 전문가로서, 대학 또는 그 이상의 전문대학원 과정에서 교육을 받고 난이도 높은 공인면허시험을 통과해야 자격을 얻을 수 있다. 단, 검안사는 안경의 조제, 가공, 판매를 할 수 없고 조제가공은 조제가공안경사만의 독자적인 영역으로 규정되어 있다.

학문의 특성이라는 면에서 볼 때, 대륙식 국가에서는 안경사가 안과 질환에 대한 진단이나 진료를 수행하는 경우가 거의 없기 때문에 교육의 내용이나 연구가 주로 시광학적인 부분에 중심을 두는 경향이 있고 학문의 명칭도 광학을 강조하는 편이다. 한국에서의 명칭이 안경광학인 점이나 독일에서의 명칭이 Augenoptik(영어로 번역하면 eye optics에 해당)인 사실이 이러한 경향성을 뒷받침한다. 그에 반해서 영미식에서는 검안학의 주요한 교육 대상이 조제가공안경사가 아닌 검안사이고, 검안사들이 안과 질환의 진단과 진료 행위를 수행하기 때문에, 교육 내용이나 연구 주제가 시광학보다는 안과학 쪽으로 발전하는 경향성이 나타난다. 검안사를 eye doctor(의사라는 의미)라고 칭한다는 사실이나, 검안학 전문대학원의 교육과정, 검안학 학술지에 실리는 논문의 분야들을 종합하여 보면 이러한 경향성을 쉽게 알 수 있다 (The Ohio State University College of Optometry 2011, University of Houston 2012).

이처럼 전문직군과 연계된 과학의 한 분야가 국가별로 학문의 내용과 직업의 특성면에서 큰 차이를 보이는 현상은 과학사적으로 매우 흥미로운 현상이라고 할 수 있다. 이처럼 국가별로 직무 영역이 크게 다른 현상은 해당 국가의 인구지리학적 요인에서 비롯되었다고 사료된다. 검안학의 발상지인 미국의 경우, 검안학이 형성되던 19세기 말부터 20세기 초까지의 기간에는

광대한 지리적 넓이에 비하여 교통 체계가 미흡하였으며, 의료보건 체계가 확립되어 있지 않고 미비한 경우가 많았다. 그 결과로 개척지인 미국 중서부(midwest)나 서부에서는 전문직인 안과의사를 접하기가 매우 어려웠고 임시방편으로 안경사가 안과의사의 역할을 수행하면서 눈 검사와 안경 처방만이 아니라 간단한 약물처방이나 진료까지 겸했던 것으로 추정된다. 이러한 안경사의 업무 범위 확장이 제도적으로 고정되면서 대륙식과는 다르게 간단한 안과 진료를 겸하는 안경사, 즉 검안사라는 독특한 직업군이 탄생하였고, 검안사들에게 필요한 안과적 지식을 교육하고 연구하기 위하여 시광학보다는 안과학 분야가 더 강조된 검안학이라는 학문이 형성된 것으로 보인다 (Goss 2003, 김재도, 이익한 2008). 이와 유사한 현상은 한국의 약사 직업군에서도 찾아볼 수 있는데, 의약분업이 확립되기 이전 한국의 약국에서는 약사가 간단한 증상의 경우 환자와의 문진을 통해 전문의약품을 처방하는 경우가 종종 있었으며, 이러한 관습이 생겼던 원인은 의사가 부족하던 근대화 시기에 약사가 의사의 역할을 대신했었기 때문이다. 한국의 약사는 의약분업 과정을 통해서 진단과 처방의 역할을 포기하고 조제와 투약에 집중한 반면에, 미국의 안경사는 안경의 조제, 가공, 판매를 포기하고 조제가공안경사에게 전담시키는 대신에 안과의사의 영역인 안과질환의 진단과 치료로 영역을 점차 확장하여 검안사로 진화해 나갔던 것으로 해석된다. 이러한 변화 과정에서 미국의 넓은 지리적 여건과 희박한 인구밀도, 그로 인하여 미흡했던 안보건 체계가 큰 영향을 미친 것으로 생각된다.

이러한 경향성은 미국과 유사하게 지리적 넓이가 넓고, 인구밀도가 희박했으며, 의료보건 체계가 미흡했던 캐나다, 호주, 뉴질랜드 등에서도 마찬가지로 나타났고, 같은 영어권이기 때문에 손쉬웠던 교류를 통하여 널리 보급되어, 현재와 같이 검안사 제도와 검안학 학문 분야가 주로 영어권 국가들에서 나타나고 영어권 국가들에 의해서 주도되는 결과를 낳았다고 할 수 있다. 같은 영어권일지라도 지리적으로 넓지 않고 인구밀도가 높았으며 일찍부터 보건의료 체계가 확립되어 있던 영국의 경우에는 법적으로 검안사의 업무가 인정받은 때가 1958년으로 매우 늦다는 사실이 이러한 검안사 제도의 진화와 인구지리학적 요인과의 관계를 간접적으로 입증한다고 할 수 있다.

1990년대 이후 최근에는 중국, 대만 등의 일부 아시아권 개발도상국이나 동유럽에서 시력 관련 의료보건 체계를 정비하면서 미국식 검안사 제도를 선진적인 체계로 간주하여 도입하는 사례도 늘어나고 있다.

반면에 독일과 프랑스 등의 대륙식 국가에서는 이러한 인구지리학적 요인이 없었고 정비된 안보건 체제 덕분에 안과의사를 쉽게 접할 수 있었기 때문에 안과의사-안경사의 2원화된 체계만으로 충분히 시력 관련 보건의료 서비스가 제공되어서, 검안사와 검안학이 안과의사나 안과

학으로부터 분화하지 않았던 것으로 해석된다. 상황이 프랑스나 독일과 유사했던 영국의 경우에 1958년에 늦게라도 검안사 제도가 도입된 것은 같은 영어권인 미국과 호주의 영향 탓이라고 이해할 수 있다.

이러한 이론을 뒷받침 하는 대표적인 예가 미국 중서부와 호주라고 할 수 있다. 미국의 중서부와 호주는 지리적으로 넓고 면적에 비하여 인구가 적기 때문에 인구밀도가 낮은 특징을 보이고 있는데, 미국에서 검안학의 교육이 가장 먼저 시작되었고 현재도 검안학의 연구와 교육이 가장 활발한 지역이 중서부라는 사실이나, 호주에서 검안학과 검안사 제도가 1920년대 부터 상대적으로 일찍 발달하기 시작하였고 현재까지도 호주의 검안학 교육과 연구의 수준이 매우 높다는 사실로부터, 안과의사로부터 분화된 검안사 직군의 발달과 그에 따른 검안학의 발달은 인구지리학적 요인과 안보건 체제의 미비함에 큰 영향을 받았음을 알 수 있다 (Goss 2003, Optometrist Association Australia 2008).

### 5. 미국에서 검안학의 발전 과정

미국은 검안학의 발상지이며, 다른 나라들의 검안학의 발전에 지대한 영향을 주었고, 현재까지도 검안학이 가장 발전된 국가이기 때문에, 미국의 검안학 발전과정을 주의깊게 살펴 볼 필요가 있다.

19세기 초반 이후 미국에서 전문적인 안경사로 활동했던 인물들은 주로 유럽에서 안경에 대한 수련을 쌓고 이민 왔던 사람들이 주류를 이루었던 것으로 추정된다 (Leasher, Pike 2009). 19세기 동안 미국의 안경사는 다른 나라와 마찬가지로 도제식으로 교육받고 전문지식을 전수하였으며 공적이고 체계적인 교육 시스템이나 법적인 면허제도 같은 것이 존재하지 않았다. 19세기 후반에 몇몇 안경 관련 교육 기관들이 설립되었으나, 오늘날로 보면 직업훈련 학원 정도의 수준이었으며 체계적인 전문 교육기관이라고는 할 수 없었다는 사실은 앞서 언급된 바 있다.

미국의 연구자들은 안과학과 독립된 검안학이 형성된 시점을 1897년으로 보고 있다. 당시 뉴욕 주의 안경사였던 C. F. Prentice가 안경 판매가 아닌 눈에 대한 검사(검안)에 대하여 고객들에게 비용을 청구하고 있었는데, 안과의사들이 이러한 행위를 불법으로 간주하여 고발하였고 안경사들이 법적으로 그에 대응했던 사건을 검안학의 시작이라고 간주하는 것이다 (Goss 2003, Leasher, Pike 2009, Fiorillo 2010). 오늘날에도 미국검안학회에서 수여하는 학술상은 Prentice의 업적을 기려서 Prentice Award라고 칭해지고 있으며, Prentice는 단순히 제도적인 면에서만 검안학에 기여했던 것이 아니라, 안경렌즈의 Prentice 법칙을 발견하여 검안학의 학

문적 발전에도 커다란 기여를 했던 인물이다. 안경사들은 당시 뉴욕 주의 법률소송에서는 승리하지 못했지만, 1901년에 미네소타 주에서 안경사의 검안 업무를 법적으로 인정한 것을 시작으로 해서 캘리포니아(1903년), 노스 다코다(1903년), 오리건(1905년), 뉴 멕시코(1906년) 등을 거쳐 1924년까지 미국의 모든 주가 안경사의 검안을 허용하게 됨으로써 최종적인 승리를 거두었다. 이러한 과정에서 앞서 언급했던 대로 동부에 비해서 인구가 희박하고 의료 서비스 체계가 상대적으로 미비했던 중부와 서부의 주들이 앞장서서 검안 업무를 법적으로 인정했다는 사실을 쉽게 알 수 있다.

당연히 이러한 안경사의 업무 영역의 확장은 안과의사들의 반발을 일으켰고, 안과의사들의 주요 반론의 근거가 안경사의 낮은 교육 수준이었기 때문에, 안경사들은 자신들의 전문 지식을 고등 교육 제도에 편입시키기를 강력하게 요청했던 것으로 보인다. 검안학/안경광학은 이렇게 고등교육을 받은 안과의사들과의 대립 과정에서 안경사들의 요구에 의하여 교육과 학문적 발전이 강화되어왔다는 점에서 상당히 독특한 면을 띄고 있다.

19세기의 직업 학원 수준이 아닌, 고등 교육으로서의 검안학이 처음 시작된 곳은 뉴욕의 컬럼비아 대학이었다. 컬럼비아 대학은 2년 기간의 단기 교육 과정으로 1910년에 검안학 교육을 시작하였으며, 이 과정은 중서부의 다른 대학 검안학 교육에 커다란 영향을 미쳤다. 그러나 컬럼비아의 검안학 교육 과정은 1953년에 대학의 결정에 의하여 폐쇄되어 버렸다. 상대적으로 인구밀도가 높고 안과진료에 접근성이 높아서 검안사의 필요성이 낮은 미국 동부의 상황에 영향을 받은 것으로 추정되며, 차후에 자세히 언급되는 것과 같이 검안학 교육이 4년제를 거쳐 5년제, 6년제로 점점 고도화 되면서 대학의 자원을 점점 더 많이 요구하게 됨에 따라 대학 측에서 경영 전략 측면에서 검안학 전공을 유지하지 않기로 결정했기 때문이라고 짐작된다.

오늘날 미국의 검안학을 주도하고 있는 연구중심대학들 가운데 가장 역사가 깊은 곳은 오하이오주립대학교(Ohio State University)와 버클리대학교(University of California, Berkeley)이다. 오하이오 주립대는 1914년, 버클리는 1923년에 검안학 교육과정을 시작하였으며 각각 미국 중부와 서부를 대표하는 유서 깊은 검안학 분야의 명문학교이다. (Fiorillo 2010, The Ohio State University College of Optometry 2011). 이 두 대학의 초기 검안학 교육과정을 살펴보면, 흥미로운 공통점이 나타난다. 한 가지는 두 대학에서 모두 검안학은 물리학과의 일부(별도의 프로그램 형식)로 출범했다는 사실이고, 또 한 가지는 교육 내용이 대부분 광학(optics)에 집중되어 있었다는 점이다. 오늘날의 모습과는 달리, 검안학은 발생 초기에는 물리학으로부터 분화된 광학의 일부라는 인식이 강했음을 알 수 있다. 오하이오 주립대의 검안학 프로그램의 초기 명칭은 응용광학(applied optics)과정이었다 (The Ohio State University College of

Optometry 2011). 또한 검안학 전공의 형성에 큰 기여를 했던 초기 교수진의 인물들도 대부분 물리학 교수들이었다. 오하이오 주립대학의 셔드(C. Sherd)나 버클리 대학의 마이너(R. S. Minor) 등이 그 예로써, 이들은 모두 물리학자들이었지만 검안학의 발전에 큰 공헌을 했던 인물들이다. 물리학과의 일부로 시작되었지만 검안학은 시간이 지남에 따라 점차적으로 학문적인 독자성을 드러내었으며 대략 2차 대전 무렵인 1940년대 초반 경에 독자적인 검안학과나 단과대학으로 독립하기 시작했다.

이렇게 물리학과 검안학이 대학 행정에서 독립적인 학과로 분화한 현상은 검안학이 광학에서 출발하기는 했지만 점차적으로 생물학, 의학의 교육과정을 첨가하면서 보건의료 학문의 성향을 띠기 시작했던 것이 중요한 원인이었던 것으로 추정된다. 검안학 교육 과정이 발생하게 된 이유 가운데 하나가 안과의사들과의 업무 영역 다툼을 위한 안경사/검안사 들의 교육과정 강화였기 때문에, 안과학 또는 그와 관련된 의학/생물학 교육의 필요성은 매우 높았을 것으로 짐작된다. 버클리 대학의 경우, 과정이 개설된 직후인 1924년에 이미 의학박사를 강사로 초빙하여 안구병리학(ocular pathology) 강좌를 개설하기 시작하였다. 흥미로운 점은 한국의 안경광학과들이 경험한 것과 유사하게 미국에서도 검안학과에서 강의하는 안과의사는 안과의사 조직으로부터 상당한 사회적 불이익을 감수해야 했다는 사실이다. 몇 가지 예들이 버클리 대학 검안학과의 역사책에 기록되어 있다 (Fiorillo 2010).

검안학이 생물학/의학적인 성격을 띠기 시작하는 이 단계에서 과학사적으로 주목해야할 점이 있다. 그것은 물리학/광학에서 시작했지만 생물학/의학을 교육과정에 강화했던 미국식 발전 방향이 필연적이었던 것이 아니라는 점이다. 호주 검안학회의 역사를 기술한 문건에 의하면, 호주에서도 미국과 유사하게 물리학에서 출발하여 생물학/의학적인 성격으로 검안학이 발전한 것은 단순히 '미국의 선례를 따랐기 때문'이라고 언급되어 있으며, 물리학/광학적인 교육에 계속 중점을 두는 것도 한 가지 대안으로 가능했던 것으로 해석하고 있다 (Optometrist Association Australia 2008). 앞서 기술했던 대로, 생물학/의학적인 교육의 강화는 업무 영역을 놓고 다투던 안과의사들과의 다툼 때문에 요구되었지만, 그것이 필연적인 발전 과정은 아니었다고 할 수 있겠다.

검안사의 업무 영역이 점차 확대되고, 안과의사들과의 영역 다툼이 강해지면서 검안학의 교육과정도 점점 길어지고 심화되는 현상을 보인다. 1940년대 검안학이 독립된 학과나 단과대학으로 분화하면서 대학원 과정이 설치되는 학교가 늘어났고 (이 때 대학원 전공의 명칭이 생리광학(physiological optics)이었다), 1950년대에는 5년제 교육과정으로, 1960년대에는 6년제 과정으로 확장되면 학위도 학사학위가 아니라 검안학 석사(Master of Optometry = M. Optom.),

검안학 박사(Doctor of Optometry = O. D.)로 수여되기 시작했다 (Fiorillo 2010, The Ohio State University College of Optometry 2011). 이처럼 교육과정이 심화되고 길어지는 현상은 기본적으로 경쟁집단인 안과의사들의 교육과정이 길어지는데 대응했던 것으로 짐작된다.

이렇게 학제가 점점 고급화 되고 더 많은 자원이 필요해지면서, 컬럼비아 대학처럼 검안학 프로그램을 접는 경우도 있었으나, 새롭게 검안학 프로그램을 시작하는 대학들의 숫자가 증가하였다. 인디애나 대학, 미주리대학(세인트 루이스), 휴스턴 대학 등의 검안학과들이 주로 이 시기에 설립되었다. 아울러 이 무렵부터 검안학과의 교수진들이 과거의 물리학자나 생물학자, 의학자 중심에서 학부부터 검안학을 교육받은 검안학자들로 충원되기 시작한다. 검안학의 발전에 크게 기여하여 오늘날까지 명성을 떨치고 있는 버클리 대학의 모건(M. W. Morgan), 오하이오 주립대학의 프라이(G. Fry), 인디애너 대학의 보리쉬(I. M. Borish) 등이 대표적인 예이다.

검안학의 역사에서 결정적인 순간은 1970년대라고 할 수 있는데, 이 기간 동안 검안학 교육과정이 미국의 의학전문대학원과 동등하게 4년제 전문대학원 과정으로 승격되었고, 1971년의 로드 아일랜드 주를 시작으로 검안사가 안과의사처럼 약물을 사용하여 눈의 질환을 진단하거나 치료할 수 있는 권한을 부여하는 법률들이 제정되기 시작하였다 (Goss 2003, Fiorillo 2010). 현재 미국의 검안사는 안경 또는 컨택트렌즈 처방을 위한 검안 업무와 더불어서, 눈의 질환을 일부 진단하거나 진료할 수 있는 권한도 가지고 있다. 본 저자의 개인적인 의견으로는, 1970년대 미국 사회에 대체 의학에 대한 관심이 급격하게 늘어났었는데, 이러한 사회적 분위기가 (비록 검안학은 대체의학이 아니고 확고하게 현대과학과 의학에 기반하고 있지만) 검안사의 진료 분야 진출에 호의적인 결과를 낳는데 일조했다고 짐작된다.

이러한 발전과정을 거치면서, 검안과 처방을 전담하는 검안사와, 처방전에 따라 안경의 조제가공만을 전담하는 조제가공안경사의 업무영역이 뚜렷하게 분화되는 현상도 나타났다. 검안학의 초기 교육과정을 보면, 안경과 렌즈의 조제가공(spectacle-making) 실습교육이 강조되고 있는데 반하여 교육과정이 고도화될수록 상대적으로 검안(eye-test)과 진단, 진료의 비중이 커짐을 알 수 있다 (Fiorillo 2010). 극도로 고도화된 검안의 필요성이 결국 검안을 안경조제가공으로부터 분리하여 독립적인 업무영역으로 분화시킨 것이다. 1954년에 이미 검안사와 조제가공안경사 집단 간의 경쟁의식에 관하여 우려하는 문건이 출판된 것을 보면, 이러한 검안과 조제가공의 분화는 1950년대 이전에 일어난 것으로 추정된다 (Koch 1954).

미국 이외의 국가에서 검안사 제도와 검안교육의 상황을 간단히 살펴 보면, 캐나다는 미국과 동일한 검안학 교육과정과 검안사 제도를 보유하고 있으며, 호주와 뉴질랜드 또한 미국과 동일한 검안사 제도를 운용하고 있으나, 교육과정은 3~5년제 학부 과정으로써 미국보다는 영국과

유사한 학제이다. 영국은 앞서 언급된대로 미국과 호주의 영향을 받아 1958년에야 검안사 제도가 도입되었으며 현재 검안학을 3년제 학부과정에서 교육하고 있다. 그 이외에 주로 과거 영국 식민지였던 영연방 국가들을 중심으로 1990년대 이후 검안사 제도와 교육과정을 도입한 나라가 30여개 국가에 이르며, 중국과 대만처럼 산업화 과정에서 미국을 벤치마킹 하면서 검안사 제도를 도입한 국가도 있다.

**6. 미국식 검안학의 발전 과정으로부터 도출되는 한국 안경광학 분야의 발전 전략**

앞서 살펴 본 검안학의 발전 과정에서 첫 번째로 눈에 띄는 특징은 검안학이라는 학문의 발전에 미친 물리학/광학의 영향력이다. 물리학의 여러 분야 가운데 광학의 연구 결과들을 사람의 눈에 적용하는 과정에서 시광학이라는 학제간 연구 분야가 탄생하였고 검안학의 태동으로 이어졌다. 또한 20세기 초반 미국에서 검안학 분야가 대학의 교육과정으로 형성되는 과정에 물리학 전공 교수들의 영향이 매우 컸음을 알 수 있다. 버클리의 마이너나 오하이오 주립대의 셰드가 모두 물리학자이면서 검안학 전공의 창설을 주도하였으며, 초기 검안학 분야의 교수진은 주로 광학을 전공하던 물리학자들이었다는 사실이 이를 뒷받침한다.

이러한 특징은 한국의 안경광학 교육 발전 과정과도 잘 일치한다. 한국의 안경광학 교육은 1980년대 전문대학 안경광학과에서 시작되었고 4년제 대학에는 1990년대 후반부터 전공학과로 설치되기 시작했는데, 주로 물리학/광학을 전공한 기존 대학 교수진에 의하여 주도되었고 현재까지도 안경광학과 교수진 가운데 물리학 전공자의 비중이 높은 편이다. 일부 안경학 전공자들은 이러한 물리학 주도의 안경광학 교육에 비판적인 입장을 견지하고 있으나, 미국의 검안학과 발전과정을 볼 때 자연스러운 현상이라고 할 수 있다. 미국에서도 학부부터 검안학으로 교육받은 인물들이 검안학과의 교수진에 합류하여 검안학의 자체 발전을 주도한 것은 20세기 중반 무렵 검안학과가 설립되고 한 세대 정도가 지난 이후부터였다. 따라서, 한국도 안경광학의 교육이 시작된지 약 30년이 지났으므로 이제 자체적인 역량을 강화시키는 것은 당연하기도 하고 또 현재 진행되고 있지만, 물리학자들에 의하여 안경광학의 초기 교육과 연구가 주도되어 왔던 사실 자체를 비판할 필요는 없다고 보여진다.

발전 과정에서 미국의 검안학과 한국의 안경광학 분야에서 나타나는 또 하나의 유사점은 안과의사 집단에 의한 강력한 견제이다. 단순히 검안사나 안경사의 업무 영역 확대에 대한 반대가 아니라, 검안학과/안경광학과에서 강의하는 안과의사에 대하여 조직적인 배타행위를 함으로써 검안사/안경사 들이 필요한 생물/의학/안과학적 지식을 습득하는 것을 어렵게 만들었다는 사실이다. 한국의 안경광학과들은 이러한 문제점을 주로 안과의사가 아닌 약학자, 생물학자,

보건학자 등을 교수요원으로 채용함으로써 극복해 왔으며, 이러한 방식은 안과의사들의 도움없이 생물/의학 교육을 실시할 수 있는 적절한 수단이었다고 평가할 수 있을 것이다.

안과의사 집단의 이러한 행동은 당연히 검안학/안경광학의 발전과 검안사/안경사의 업무 범위 확장이 안과의사들의 업무 영역을 침해할 것이라는 우려 때문이다. 이러한 행동은 단순히 교육에 대한 방해에만 국한되지 않으며, 교육을 바탕으로 한 검안사/안경사들의 업무 범위 확장 노력에 대한 집단적인 저항으로 나타나고 있다. 현재 한국에서도 검안에 필요한 약물 점안 문제, 콘택트렌즈 처방을 위한 세극등 사용 문제, 간단한 안압 측정이나 시야 검사 등을 통한 녹내장 조기 진단 문제 등 더 정확한 시력검사를 위한 안경사의 업무 범위 확장이 안과의사 집단의 저항에 의하여 지연되고 있다.

이러한 현실에서, 미국식 검안학의 발전 과정과 비교하여 한국의 안경광학은 독자적인 발전 방향을 설정할 필요가 있다고 본다. 안과의사의 업무 영역에 대한 침해를 최소화하여 안과의사들의 집단 저항을 감소시키면서 안과의사와 중복되지 않는 안경사만의 독특한 전문성을 강조한다면, 정확한 시력교정을 위하여 필요한 업무 범위 확장을 획득하기가 상대적으로 용이하리라고 생각된다. 이 문제는 단순히 업무 범위만의 문제가 아니라, 한국의 안보건 시스템이 영미식 안과의사-검안사-조제가공안경사 3원화 체제로 진화해야 하는가, 아니면 대륙식 안과의사-안경사 2원화 체제로 존속해야 하는가 하는 선택의 문제와도 관련이 있다는 것이 본 저자의 의견이다. 왜냐하면, 영미식 검안사의 경우에는 안경의 조제가공을 수행하지 않고 안경 또는 콘택트렌즈 처방에만 수입을 의존하기 때문에 그 업무 영역이 매우 좁아져서 필연적으로 안과 질환의 진단이나 진료 같은 안과의사의 업무 영역으로 진출하기를 모색할 수밖에 없기 때문이다. 이러한 가능성이 안과의사 집단의 강력한 반발을 더욱 촉진하는 악영향을 미치고 있다. 뒤에 논의하는 바와 같이, 영미식 안과의사-검안사-조제가공안경사 3원 체제가 무조건 바람직한 보건의료 시스템이라고 할 수는 없으며 한국의 실정과도 맞지 않으므로, 한국의 안경광학계는 미국식 검안사 모델의 도입을 추구하면서 안과의사 집단과 충돌을 일으켜 안경사에게 필요한 검안 업무에 마저 제약을 받게 되는 부적절한 전략보다는, 안질환의 진단과 진료 분야에서는 안과의사의 독자성을 인정하고 안경사의 전문성은 눈의 광학적인 측면에 두어서 그에 꼭 필요한 검안 업무(예를 들면 굴절 검사에 있어서 조절 마비제의 점안, 세극등의 사용 등)를 가능하게 하도록 노력하는 방향으로 발전해 나가는 것이 적절하다고 여겨진다. 앞서 설명한 대로 안과의사와 안경사의 장점은 각각 안질환과 시광학으로 서로 다르기 때문에, 영미식 검안사 제도의 경우에 나타났던 것과 같은 업무 영역의 중복과 경쟁 보다는, 각자의 전문 영역을 구분하여 집중하는 것이 서로에게 더 이익이라는 점을 고려할 필요가 있다.

같은 맥락에서 영미식 안과의사-검안사-조제가공안경사의 3원화된 안보건 체제의 도입은 한국의 안경광학 발전에 부정적인 영향을 미칠 것이라고 본 저자는 생각한다.

우선, 사회 구조적인 측면에서 보았을 때, 한국은 일제 강점기를 거치면서 사회 시스템이나 구조가 기본적으로 대륙식 구조를 가지게 되었다는 점을 지적하고자 한다. 일본은 명치유신 개화기에 주로 독일을 벤치마킹했기 때문에 대륙식 사회 시스템을 가지고 있고, 일제 강점기 동안 한국의 근대화 과정이 일본의 영향을 크게 받은 관계로 한국 또한 사회 시스템에서 대륙식의 특징이 강하게 나타난다. 해방 이후 미국의 영향을 받아 영미식 제도를 한국에 수입하려는 노력이 계속되어 왔지만, 대륙식 사회 시스템의 기반 위에 영미식 제도의 말단만 수입해서 이식했기 때문에 성공하지 못한 경우가 매우 많았다. 대표적인 예가 교육 제도이다. 미국식 로스쿨이나 의학전문대학원 제도를 실시하였으나, 기본적으로 독일식인 한국의 교육 제도와 부조화를 일으켜서 결국 의학전문대학원은 의과대학으로 환원되고 로스쿨 또한 많은 문제점을 노출하고 있다. 대학의 전공 선택 방법 또한 입학 전에 전공을 미리 결정하는 유럽식을 배제하고, 미국식 자유 전공 선택 제도를 도입하고자 수차에 걸쳐 시도하였으나 모두 실패한 바 있다. 유럽식 사회 시스템에 단순히 영미식 제도 일부만 이식하는 것은 성공하기 어려움을 알 수 있다. 안과의사-검안사-조제가공안경사 3중 체제는 명백하게 영미식의 독특한 제도로서, 교육제도 등과 마찬가지 이유로 대륙식 사회 시스템인 한국 사회에 이식하기 매우 어렵다는 것이 본 저자의 의견이다.

검안사 제도 도입의 어려움은 단순히 사회 시스템 문제에만 국한되지 않는다. 앞서 설명했던 대로, 검안사 제도의 도입과 발전은 지리적 광범위성, 희박한 인구밀도 등과 같은 인구지리학적 요인과 그에 따른 안보건 체제의 미비함이 주요한 원동력이었다. 검안사의 업무 범위가 넓어진다 해도 안과의사와 동등한 업무를 수행할 수 없는 현실에서 검안사의 존재 가치를 사회적으로 인정받기 위해서는 안과의사의 희소성이 절대적으로 요구되기 때문이다. 한국에서 검안사 제도의 도입이 1960년대 이전에 시도되었다면 가능성이 있었을지도 모른다. 그러나 21세기 현재 한국은 충분히 선진화된 보건의료체계를 갖추고 있으며, 교통이나 기타의 문제로 안과의사에 대한 접근성이 현격하게 떨어지는 지역은 거의 존재하지 않는다. 이미 안과의사만으로도 안질환의 진단과 진료에 필요한 인력이 충분히 공급되고 있다 (안과의사들은 공급과잉이라고 주장하고 있다). 그러므로 한국은 안과의사와는 별도로 검안사 제도를 도입할 필요가 있는 단계를 이미 넘어섰으며, 검안사 제도의 도입 주장은 그 필요성 측면에서나 효율성 측면에서 부정적으로 평가될 수밖에 없다고 사료된다.

현재 한국의 안경광학계 일부에서는 검안사 제도의 도입을 적극 주장하면서 커다란 기대를

걸고 있거나, 수년 내에 검안사 제도가 도입될 것으로 확신하면서 유사한 주장을 담은 문건을 계속 공개하는 경우가 있다. 검안사 제도의 도입이 계속 주장되는 이유는 대략 두 가지 때문이라고 본 저자는 추정하고 있다.

첫 번째 이유는 안경광학 분야에서 나타난 학제간 불일치 현상 때문이다. 초기 단계에서 전문대학 교육으로 출발한 안경광학 전공이 1990년대 후반부터 4년제 대학에 개설되기 시작하면서 전문대학과 4년제 대학 사이에 이해관계가 일치하지 않는 현상이 자주 나타났다. 검안사 제도의 도입 주장은 이 시기에 전문대학 안경광학과 측에서, 4년제 대학은 검안사 교육을 담당하고 전문대학은 조제가공안경사 교육을 담당하자는, 분업화의 논리로 개발된 사정이 있다고 추정된다. 직설적으로 표현하면 4년제 대학 안경광학과 개설에 따른 전문대학 안경광학과의 생존논리로서 검안사 제도의 도입이 주장된 것이다. 2012년 현재 시점에서는, 여러 전문대학 안경광학과가 4년제로 승격하거나 3년제로 전환하면서 자체적인 학사학위 취득 제도를 도입하여 과거의 4년제 대학과 전문대학 간의 마찰이 상당 부분 감소하였고, 학제간 분업화에 기인한 검안사 제도의 도입 주장은 상당히 감소한 상황이다.

두 번째 이유는 일부 유학파의 해외 제도 수입 주장 때문이다. 해외 유학생들에게 종종 나타나는 현상 가운데 하나는 자신이 유학한 나라의 문물을 무조건적으로 칭송하면서 역사.문화.사회적인 고찰 없이 곧바로 한국에 이식해야 한다는 주장하는 것이다. 본인들이 직접 의식하지는 못하지만, 이러한 주장은 해당 국가의 제도를 선진적인 것으로 높이 평가하면서 은연중에 그 나라에서 공부한 자신의 가치를 높이고자 하는 잠재의식도 포함되어 있다. 이러한 이유로 인해서 초기에 해외로 유학하여 안경광학을 공부하고 귀국한 일부 유학파 안경광학자들 가운데는 미국과 호주에서 검안사의 사회적 위상과 전문성을 지나치게 높이 평가하고 검안사 제도를 도입해야 한다는 주장이 강했었다. 그러나 여러 자료를 통하여 조사된 바에 따르면, 검안사 제도가 가장 발달한 미국, 캐나다, 호주에서조차도 검안사의 사회적 위상은 한국의 기대만큼 높지 않으며, 검안사가 무슨 일을 하는 사람인지 해당 협회에서 광고 방송을 계속해야 할 정도로 사회적 인식도 널리 퍼져 있지 못하고, 대부분의 환자들은 검안사보다는 안과의사를 더 선호하는 것으로 보인다.

검안사 제도의 도입을 추구하는 일부의 움직임 때문에 안과의사 집단과의 마찰이 더욱 심화되어, 검안사 제도의 도입은 커녕, 한국의 안경사 집단이나 안경광학계는 안경사의 굴절검사 업무에 필수적인 업무 영역의 확대조차도 애로를 겪는 상황이 초래되는 결과를 낳았다고도 볼 수 있다.

한국의 안경사 집단에서 오해하고 사항들 가운데는 미국과 호주의 검안사의 높은 수입도 있

다. 미국이나 호주의 검안사가 안과의사에 버금가는 수준의 교육을 받는 전문직이기 때문에 일반 직업군에 비하여 고소득을 올리는 것이 사실이다. 그러나 검안사의 이러한 고소득이 가능한 이유는 안경조제가공 이외에도 검안 업무에 높은 비용을 책정하는 영미권의 제도적 특징 때문이다. 안경이나 콘택트렌즈를 맞추기 위하여 기본 검안을 1회 실시할 때, 나라별 지역별 격차가 있기는 하지만, 미국의 경우 그 비용이 약 70 달러(한화 8만원 상당)에 달하는 것으로 알려져 있다. 아직까지도 검안이 무료이거나 (안경사) 적은 액수의 기본 의료서비스로 간주되는 (안과) 한국에서, 검안사 제도가 정착하려면 미국에 준할 정도로 높은 검안 비용이 책정되어야 한다는 점을 알 수 있다. 제한된 보건의료 재원에 허덕이고 있는 현실과, 안과의사-안경사 2원화 체제만으로도 영미식 검안사의 모든 업무를 수행할 수 있다는 점을 고려해 볼 때, 단순히 안경사 집단의 이익만을 위하여 검안사 제도를 도입하자고 주장하는 것은 대중적 설득력이 없다고 생각된다. 한국에서 검안사 제도의 도입은 안과의사 집단의 반대만 극복하면 가능한 것이 아니라, 검안사 집단의 경제적 소득을 뒷받침할 보건의료 재원의 확보가 이루어지지 않으면 불가능할 것이며, 그 재원의 확보를 위한 검안 비용의 대폭적인 증가는 현실적으로 가능성이 높지 않다. 검안사 제도의 도입을 주장하는 일부 집단에서는 미국과 호주의 검안사는 안경과 콘택트렌즈 처방을 위한 검안만이 아니라 안질환의 진단과 치료로도 수입을 올리고 있다는 주장을 펴기도 한다. 그러나 미국 인디애너 대학의 졸업생 설문 조사에 따르면 검안사의 수익 가운데 안질환에 대한 진단과 진료로부터 얻는 수입은 7%에 불과한 것으로 나타나서 (Goss, Grosvner 1998), 검안사의 수입 가운데 대부분은 안경과 콘택트렌즈 처방을 위한 검안으로부터 나오고 있음을 쉽게 알 수 있다. 한국에 검안사 제도가 도입된다 하더라도, 대부분의 고객들은 안질환 진단과 진료를 위하여 검안사보다는 안과의사를 선호할 것이며, 검안사의 수입은 주로 굴절검사에 의존하게 될 것으로 예상함이 타당하다. 그리고 검안 비용이 현격히 낮은 한국의 현실에서 볼 때 검안사는 전문직에 걸맞는 수익을 올리기가 매우 어려울 것이고 자칫 한약사 제도처럼 유명무실한 전문면허로 추락할 가능성이 매우 높다.

이러한 이유들로 인해서 본 저자는, 일부 한국 안경계의 주장과는 달리, 영미식 검안사 제도의 도입은 한국 안경광학의 발전에 긍정적인 효과보다는 부정적인 면이 더 많다고 생각한다.

그렇다면 한국 안경광학의 바람직한 발전 방향은 무엇일까? 앞서 언급한 역사.문화.사회적 맥락에서 고찰해 볼 때, 한국 안보건 체제의 원형인 독일식 발전 모델을 추구하는 것이 가장 적절하다고 본다. 독일식 모델은 대륙식이기 때문에 한국의 사회 시스템과도 잘 일치하고, 또 현재 한국 안경광학계의 현실과도 정합성이 뛰어나다. 먼저 안경사 교육은 현재와 같이 대학 또는 전문대학에서 시행한다. 독일에서 안경사는 기술대학(Fachhochschule) 수준에서 교육되

고 있다. 일부에서는 기술대학을 공업고등학교로 잘못 번역하기도 하지만, 기술대학은 독일에서 중등교육을 이수한 후에 (고교 졸업 후에) 진학하여 4년간 교육 받는 고등교육기관으로서 한국의 대학/전문대학과 같은 수준에 해당한다. 기술대학에서는 졸업 후 안경광학을 더욱 심도 있게 공부하여 한국의 석사학위에 상응하는 학위도 취득할 수 있다. 교육받은 안경사는 한국과 마찬가지로 굴절검사, 검안, 안경조제가공, 판매를 수행할 수 있으며 안과의사와 함께 안보건 체계를 분담하고 있고 별도의 검안사 제도는 존재하지 않는다. 물론 콘택트렌즈도 어무범위에 포함되어 있다. 독일의 안경사는 오랜 노력을 통하여 한국의 안경사 보다 검안 업무의 범위가 넓어져 있으며, 일정 정도 약물 점안을 통한 검안이나 시야 검사를 통한 녹내장 조기 진단도 가능한 수준에 이르러 있다.

한국의 안경광학 발전을 위해서는, 한국 실정에 부합하지 않는 검안사 제도의 도입을 추진하여 안과의사 집단과의 마찰만 증폭시키고 업무 범위 확대를 답보 상태로 만드는 것보다, 안과의사-안경사로 구분된 대륙식 시스템을 확고하게 인정하고, 안과의사는 안질환의 진단과 진료에 안경사는 시광학과 광학적 시력 교정에 전문성이 있음을 서로 인정한 다음, 서로의 전문성에 대한 침해를 최소화 하면서 각각의 전문성을 극대화하기 위하여 필수적인 업무 영역 확대 (검안을 위한 약물 점안이나 세극등의 사용)에만 역량을 집중하는 것이 가장 바람직하다고 생각된다. 이러한 노력이 관철된다면 한국의 안경사는 눈의 광학적인 시력 교정 분야의 독보적인 전문가로서 굴절검사에서부터 콘택트렌즈를 거쳐 안경조제가공까지를 담당하는 고소득 집단으로 정체성을 확보할 수 있으리라 기대된다.

검안학/안경광학은 해당 전문직업과 밀접하게 연계된 학문이기 때문에, 전문직업의 발전 방향이나 업무 범위에 따라서 학문 자체의 발전 방향과 연구 내용이 크게 영향을 받을 수밖에 없다. 미국식 검안학을 모델로 발전 방향을 설정할 경우에는 앞서 언급된 대로 학문의 내용이 광학 분야로부터 시작하여 생물/의학 분야의 연구로 무게중심이 이동하게 되는 것이 보통이다. 검안사의 경쟁 상대인 안과의사 집단에 대응하기 위해서 그와 동등한 교육이 필요해지고 안과학이 교육 및 연구의 핵심으로 자리하게 되는 것이다. 반면에 독일식 안경광학 모델로 발전 방향을 설정할 경우에는, 질환 중심인 안과학적 연구보다는 시광학을 중심으로 광학, 생물의학, 안경학의 균형있는 교육과 연구가 가능해진다. 이러한 경향성은 출판되는 논문의 주제들로부터 알 수 있는데, 미국검안학회지(Optometry and Vision Science)나 호주검안학회지(Clinical and Experimental Optometry)가 주로 의학 저널과 유사한 체제를 가지고 있고 게재되는 논문의 연구주제가 생물학/의학 관련 비율이 압도적으로 높은데 반하여, 유럽에서 출간되는 Opthalmic and Physiological Optics 저널이나 Vision Research 저널은 시광학과 관련된 주제

의 논문을 주로 게재한다는 점이 흥미롭다. 호주검안학회의 문건에서 언급된 대로, 검안학/안경광학이 발전함에 있어서 안과학적인 방향으로의 강조와 전환은 필수적인 것이 아니며 선택 가능성들 가운데 하나일 뿐이라고 할 수 있다. 한국의 안경광학계는 미국식 검안사 모델이 아니라 독일식 안경사 모델을 추구해야 하며, 안경광학 교육과 연구에 있어서 물리학, 생물의학, 안경학이 고르게 균형 잡힌 종합학문으로 발전해 나가는 것이 가장 바람직한 방향이라고 본 저자는 생각한다.

## 7. 요약 및 결론

검안학/안경광학의 발전과정을 과학사적인 측면에서 고찰하고 그로부터 한국 안경광학의 바람직한 발전 방향과 전략을 도출하였다. 과학사적으로 검안학/안경광학은 물리학/광학의 전통에서 출발하여 생물의학과의 학제적 연구를 통하여 시과학/생리광학의 단계를 거쳐 검안학/안경광학으로 발전하였으며, 이 과정에서 19세기에 보기 드문 학제적 연구와 의학생물물리학의 특성을 가졌음을 알게 되었다. 검안학/안경광학이 광학이나 안과학으로부터 독립하여 독자적인 학문과 직업군으로 형성되는 과정에서 인구지리학적 요인이 커다란 영향을 미쳤음을 알 수 있었다. 이러한 고찰로부터 한국의 현실에 적절한 발전 모델은 미국식 검안학이 아닌 독일식 안경광학임을 도출하였으며, 이러한 연구결과는 한국 안경광학계의 발전 전략 수립에 기여할 수 있으리라 기대된다.

## 8. 감사의 글

# 참고문헌